\documentclass[twocolumn,prl,aps,showpacs]{revtex4}
\usepackage{graphicx}

\begin{document}

\title{\bf Adsorption Transition of a Polyelectrolyte on a High-dielectric Charged
Substrate}

\author{Chi-Ho Cheng}
\email{phcch@hotmail.com}
\author{Pik-Yin Lai}

\affiliation{Department of Physics and Center for Complex Systems,
National Central University, Taiwan 320, Republic of China.}

\date{\today}

\begin{abstract}
The behavior of a polyelectrolyte adsorbed on a charged surface of
high-dielectric constant is studied by both Monte-Carlo simulation
and analytical methods. It is found that in a low ionic strength
medium, the transition is first-order with the repulsive charged
surface. The surface monomer density, which is the order parameter
of the adsorption transition, follows a linear relation with
surface charge density. It indicates the polyelectrolyte is
compressed on the substrate without any conformational change
before the de-sorption. Finally, a new scaling law for the layer
thickness is derived and verified by simulation.
\end{abstract}

\pacs{61.25.Hq, 82.35.Gh}

\maketitle

\vspace{-5pt}


Polymer adsorption on an attractive surface has drawn considerable
interest due to its relation to surface effects in critical
phenomena and practical importance in material science and
biophysics. It is well established that the adsorption transition
is continuous if its attraction on the surface is short-ranged
\cite{degennes,degennes2,eisen82,lai}. On the other hand,
long-ranged electrostatic interactions in polyelectrolyte systems
pose many challenging theoretical problems. Recently the macroion
adsorption on an electrostatically attractive interface and the
associated charge inversion phenomena of adsorbed polyelectrolytes
acquire lots of attention \cite{gelbart,book,review2}.

Previous analytical approaches using the Edwards equation imposed
the continuity of the monomer density across the surface and
setting the monomer density to zero at the surface
\cite{wiegel,muth}. Within the framework of the self-consistent
field method, both the Poisson-Boltzmann equation and Edwards
equation were solved simultaneously
\cite{varoqui,borukhov,chatellier} with zero monomer density at
the surface. These treatment however cannot faithfully respect the
electrostatic boundary condition. As a result, the adsorption
transition would still be continuous whatever the surface
potential looks like.

Recently, the theoretical interest to the problem is due to its
importance for multi-layer polyelectrolyte adsorption
\cite{joanny,netz,solis}. It also raises the question of applying
Poisson-Boltzmann theory to polyelectrolyte adsorption because the
theory fails to capture the correlation effects.

In this letter, we study the adsorption of a single
polyelectrolyte on a high-dielectric substrate in which the image
charge attraction is strong. At low ionic strength, the adsorption
transition occurs when the surface charges are repulsive instead
of the attractive case that were usually studied. The problem is
tackled by performing Monte-Carlo simulations and also by
analytical methods in polymer physics taking full account of the
appropriate boundary conditions. It is found that the order of the
adsorption transition, the physical mechanism, and the scaling
behavior are all different from those of the attractive surfaces.


A polyelectrolyte carrying positive charges is immersed in a
medium ($z>0$) of dielectric constant $\epsilon$. At $z=0$ there
is an impenetrable surface of uniform surface charge density
$\sigma$. Below the surface ($z<0$), it is a substrate of
dielectric constant $\epsilon'$. Denote the charge on a polymer
segment $ds$ by $q_0 ds$, the Hamiltonian is
\begin{widetext}
\begin{eqnarray}
{\cal H} &=&  \frac{3k_{\rm B}T}{2l_0^2} \int_0^N ds
\left(\frac{\partial{\vec r(s)}}{\partial{s}}\right)^2 +
\frac{1}{2} \int_0^N ds \int_0^N ds' \left( \Gamma\frac{{\rm
e}^{-\kappa|\vec r(s) - \vec r(s')|}}{|\vec r(s) - \vec r(s')|} -
\Gamma'(2-\delta_{s,s'})\frac{{\rm e}^{-\kappa|\vec r(s) - \vec
r'(s')|}}{|\vec r(s) - \vec
r'(s')|} \right) \nonumber \\
&&+ h \int_0^N ds \kappa^{-1}{\rm e}^{-\kappa \vec r(s)\cdot\hat
z} + \omega\int_0^N ds\int_0^Nds' \delta(\vec r(s)-\vec r(s'))
\label{ham}
\end{eqnarray}
\end{widetext}
where $s$ is the variable to parametrize the chain, $l_0$ the bare
persistence length, and $\kappa^{-1}$ the Debye screening length.
$\vec r(s)=(x(s),y(s),z(s))$, $\vec r'(s')=(x(s'),y(s'),-z(s'))$
are the positions of the monomers and their electrostatic images,
respectively.  $\Gamma = q_0^2/\epsilon$, $\Gamma' =
\Gamma(\epsilon'-\epsilon)/(\epsilon'+\epsilon)$, and $h=4\pi
q_0\sigma/(\epsilon'+\epsilon)$ are the coupling parameters
governing the strength of Coulomb interactions among the monomers
themselves, between the polymer and its image, and between the
polymer and the charged surface, respectively. The last term in
Eq.(\ref{ham}) represents the excluded volume interactions with
$\omega>0$ (good solvent regime) in this study. We shall focus on
the case of a charge polymer in a low ionic strength medium.


The above continuum model is discretized to perform Monte-Carlo
simulation. The continuous curve $\vec r(s)$ is replaced by a
chain of beads $\vec r_i$ ($i=1,\ldots,N$) with hard-core excluded
volume of finite radius $a$. Total lengths up to $N=120$ are
employed. Units of length and energy are set to be $2a$ and
$q_0^2/2\epsilon a$, respectively. Dielectric ratios
$\epsilon'/\epsilon$ are chosen from 2 to 12.5 (aqueous solution
with a metallic substrate). Runs up to $10^9$ MC steps are
performed to achieve good statistics.

\begin{figure}[tbh]
\begin{center}
\includegraphics[width=3in]{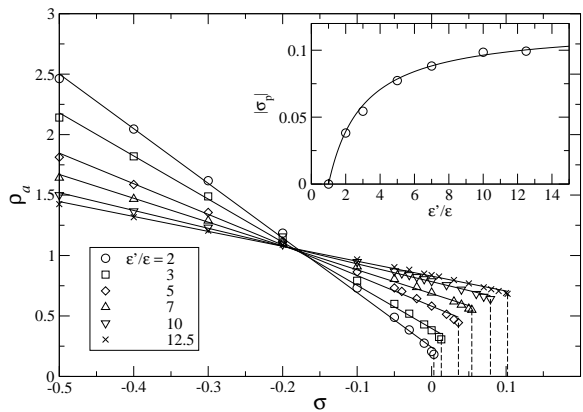}
\end{center}
\vspace{-5pt}
 \caption{Monte-Carlo results for the normalized monomer density at the surface,
$\rho_a$, as a function of surface charge density $\sigma$ (in
unit of $q_0/4a^2$) for different $\epsilon'/\epsilon$ at
$\kappa^{-1}=25$. The fitted straight lines are terminated at
their adsorption transition points. The vertical dashed lines are
drawn as guides to the eyes. Inset: The polarization surface
charge density induced by the polyelectrolyte, $|\sigma_{\rm p}|$,
as a function of dielectric ratio $\epsilon'/\epsilon$. The sign
of $\sigma_{\rm p}$ is opposite to $q_0$ and is negative. The
solid curve is fitted from Eq.(\ref{sigmapoly}) with $\sigma_{\rm
poly}=0.118$. }
 \label{adsorb6.eps} \vspace{-5pt}
\end{figure}

The adsorption layer can be characterized by the normalized
monomer density $\rho(z)$. $\rho_a\equiv\rho(a)$ representing the
fraction of monomers being adsorbed on the substrate is chosen as
an order parameter to describe the adsorption transition.
$\rho_a>0$ and $\rho_a=0$ characterize the adsorbed and de-sorbed
states respectively. In Fig.\ref{adsorb6.eps}, $\rho_a$  as a
function of the surface charge density $\sigma$ for various
$\epsilon'/\epsilon>1$ is shown. The discontinuous jump of
$\rho_a$ across the threshold indicates the transition is first
order. We also verified that the energy jump (latent heat) across
the transition is proportional to $N$. Similar results were
obtained for larger $\kappa^{-1}$.

\vspace{15pt}

Furthermore, the data in Fig.\ref{adsorb6.eps} also indicate that
$\rho_a$ is linear in $\sigma$ with the slope depending on the
ratio of $\epsilon'/\epsilon$. Such a linear relation between
$\rho_a$ and $\sigma$ can be understood from the electrostatic
boundary conditions that the system has to satisfy. The electric
potential $\phi(z)$ in the neighborhood of the $z=0$ boundary
obeys,
\begin{equation}
-\left.\frac{\partial \phi}{\partial z} \right|_{z=0^+} +
 \left.\frac{\partial \phi}{\partial z}\right|_{z=0^-}
 = -\frac{4\pi}{\epsilon}\left(\frac{2\sigma}{\epsilon'/\epsilon+1}+\sigma_{\rm p}\right)
 \label{boundary1}
\end{equation}
where $\sigma_{\rm p}$ is the polarization surface charge density
induced by the polymer only, which depends on $\epsilon'/\epsilon$
but is independent of $\sigma$ in the adsorbed regime near the
transition. Notice that $\sigma_{\rm p}$ in general is a
complicated function since it relies on the polymer conformation.
Also, if one treats the polymer as a marcomolecule with a
well-defined surface, its surface charge density at $z=a$ should
be proportional to the monomer density $\rho_a$, this also applies
to the electric field in the $z<0$ region,
\begin{eqnarray}
K \left.\frac{\partial \phi}{\partial z}\right|_{z=a^-} &=&
-\frac{4\pi}{\epsilon}\rho_a , \\ K \left.\frac{\partial \phi}
{\partial z}\right|_{z=0^-}  &=&
-\frac{4\pi}{\epsilon}\frac{2\epsilon'}{\epsilon'+\epsilon}\rho_a
 \label{boundary2}
\end{eqnarray}
where $K>0$ is the corresponding proportional constant. Applying
the continuity condition for the electric field from $z=0^+$ to
$z=a^-$, and using Eqs.(\ref{boundary1})-(\ref{boundary2}), one
gets the linear behavior
\begin{equation}
\rho_a = -\frac{2K}{\epsilon'/\epsilon-1} \left(\sigma +
\frac{\epsilon'/\epsilon+1}{2}\sigma_{\rm p} \right)
\label{rho-sigma}.
\end{equation}
Notice that $K$ and $\sigma_{\rm p}$ are functions of
$\epsilon'/\epsilon$. The linear behavior in Eq. (\ref{rho-sigma})
is confirmed by the simulation data as shown in
Fig.\ref{adsorb6.eps}, the slope decreases monotonically with
$\epsilon'/\epsilon$. Substituting $\sigma=0$ into
Eq.(\ref{rho-sigma}), we get the polarization surface charge
density as a function of dielectric constant ratio,
\begin{equation}
\sigma_{\rm p}= -\frac{\rho_a|_{\sigma=0}}{K}
\frac{\epsilon'-\epsilon}{\epsilon'+\epsilon}. \label{sigmapoly}
\end{equation}
$K$ for different dielectric ratios is obtained from the slopes of
different straight lines presented in Fig.1. $\sigma_{\rm p}$ as a
function of $\epsilon'/\epsilon$ is then fully determined from
simulation data as shown in the inset of Fig.1. It fits very well
to Eq.(\ref{sigmapoly}) with $\sigma_{\rm poly}\equiv
\rho_a|_{\sigma=0}/K=0.118$. It suggests the conformation of the
adsorbed polymer is compressed on the high-dielectric substrate.
We have also check the adsorption layer thickness is independent
of the number of monomer $N$, which is consistent with the picture
of a compressed state. There is no conformational change during
the de-sorption contrary to the scaling results as predicted by
Borisov {\it et.al.} \cite{borisov} for the attractive surface.


The polyelectrolyte behaves as electric blobs arranged
longitudinally and lie down parallel to the surface. Increasing
the attraction from the charged surface reduces the
$z$-fluctuation amplitude of the chain, but the effective in-plane
surface charge distribution of the polyelectrolyte does not
change. Hence the polarization $\sigma_{\rm p}$ is independent of
$\sigma$. The excluded volume effect is safely ignored because it
takes almost no effect in the $z$-direction. The effect from
self-electrostatic interaction of the polyelectrolyte can be
absorbed into the bare persistence length from $l_0$ to $l$.

Because the monomer would feel the strongest attraction from its
direct image around the adsorption regime, the $\Gamma'$-term in
Eq.(\ref{ham}) is approximated by the interaction of each monomer
and its corresponding image only. The residual attraction from the
images of other monomers is then adsorbed into the coupling
parameter $\Gamma'$ from $q_0$ to $q$. The partition function is
reduced to
\begin{eqnarray} \label{partit}
Z = \int{\cal D}[\vec r(s)]\exp  [\int_0^N ds \{-\frac{3}{2l^2}
 \left(\frac{\partial{\vec r(s)}}{\partial{s}}\right)^2
\nonumber \\  + \frac{\beta\Gamma'}{4}
 \frac{{\rm e}^{-2\kappa \vec r(s)\cdot\hat z}}{\vec r(s)\cdot\hat z}
 - \beta h \kappa^{-1} {\rm e}^{-\kappa \vec r(s)\cdot\hat z}
 \}].
\end{eqnarray}
Transforming the variable from $\vec r(s)$ to the normalized
monomer density $\rho(\vec r)=\frac{1}{N}\int_0^Nds \delta(\vec
r-\vec r(s))$ by introducing an auxiliary field, then applying the
ground state dominance approximation in large-$N$ limit and by
variational principle \cite{doi,orland}, one obtains the
Edwards-Schr\"{o}dinger equation,
\begin{equation} \label{sch}
\left(-\frac{l^2}{6}\frac{d^2}{dz^2} -\frac{\beta\Gamma'}{4}
\frac{{\rm e}^{-2\kappa z}}{z} + \beta h \kappa^{-1} {\rm
e}^{-\kappa z} \right) \psi(z) = \varepsilon_0 \psi(z)
\label{tise}
\end{equation}
where $\varepsilon_0$ acts as a Lagrange multiplier to enforce the
constraint of the ground state wavefunction normalization. The
monomer density is given by $\rho(z)=|\psi(z)|^2$. Eq.(\ref{tise})
also describes a quantum particle at its ground state moving under
a combined potential of a 1-d screened Coulomb attraction and an
almost linear potential. However, the boundary condition expressed
by Eq.(\ref{rho-sigma}) is different from the hard-wall boundary
condition $\psi|_{\rm s}=0$ usually employed for a quantum
particle. Instead $\psi|_{\rm s} = \sqrt{\rho_a} \neq 0$ for the
present problem implies that the steric force felt by the
polyelectrolyte from the charged surface should be modified
\cite{steric}. Setting $\psi|_{\rm s} = 0$
\cite{wiegel,muth,review1} in the polyelectrolyte adsorption
problems in previous studies are not completely correct.

During the adsorption, the rod-like polyelectrolyte tends to lie
down on the charged surface. The thickness of the adsorption layer
is of the same order of the gyration radius in z-direction. At low
ionic strength in which the Debye length is much greater than the
layer thickness, the polyelectrolyte cannot feel the potential of
length scale much larger than $\kappa^{-1}$, but only the
potential barrier height is important. The original potential
$V(z)$ in Eq.(\ref{sch}) can thus be replaced by
\begin{equation} \label{potential}
V_{{\rm mod}}(z) = \cases{  +\infty , &$\qquad z < a $ \cr V(z) ,
&$\qquad a \leq z < z_{\rm br}   $\cr  V(z_{\rm br}), &$\qquad  z
\geq z_{\rm br} $\cr }
\end{equation}
where $z_{\rm br}$ is chosen such that $V'(z_{\rm br})=0$ and
$V(z_{\rm br})$ is the barrier height. In the limit of $\sigma =
\kappa=0$, analytic solution gives
\begin{equation}
\psi(z) = {\rm
W}_{\lambda,1/2}(\frac{3\beta\Gamma'}{2l^2\lambda}z)
 \hspace{1mm};\hspace{1mm}
\varepsilon_0 =
-\frac{3\beta^2\Gamma'^2}{32l^2}\frac{1}{\lambda^2}
\label{binderg}
\end{equation}
where ${\rm W}_{\lambda,1/2}$ is the Whittaker's notation of the
confluent hypergeometric function \cite{1dcoulomb}, and $\lambda$
is the least value satisfying the boundary condition. Bound state
exists for arbitrary $\epsilon'/\epsilon > 1$. It implies the
threshold surface charge density, $\sigma_{\rm t} > 0$ at low
ionic strength.

\begin{figure}[tbh]
\begin{center}
\includegraphics[width=3in]{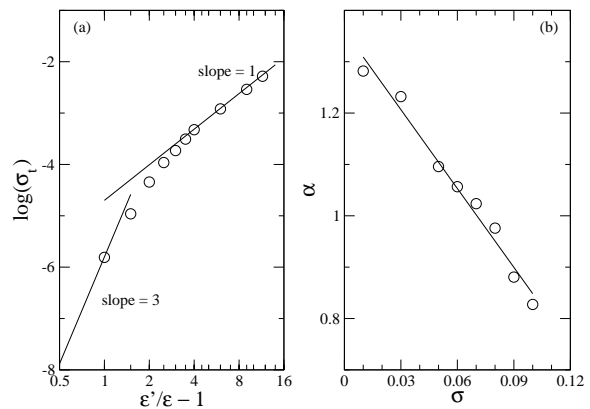}
\end{center}
\vspace{-5pt}
 \caption{
(a) Simulation results for the surface charge density at the
transition, $\sigma_{\rm t}$, as a function of dielectric ratio
$\epsilon'/\epsilon$ in logarithmic scale at $\kappa^{-1}=25$. The
straight lines indicate slopes of 1 and 3 as suggested in the
text. (b) Simulation results for the inverse decay length $\alpha$
(which is proportional to the inverse layer thickness) as a
function of $\sigma$ (in unit of $q_0/4a^2$) for
$\epsilon'/\epsilon=12.5$, $\kappa^{-1}=25$. The straight line is
a linear fit. $\sigma_{\rm t}=0.102$ in this case.  $\alpha$ is
obtained from exponential fitting to the tail of corresponding
density profile.}
 \label{sigma3.eps}
 \vspace{-5pt}
\end{figure}

For both $\sigma, \kappa > 0$, no exact solution exists in general
but one can analyze it around the transition. Near the surface,
the image charge attraction dominates over the surface charge
repulsion and hence the binding energy is approximated by the
$\sigma = 0$ case in Eq.(\ref{binderg}). The polyelectrolyte
undergoes a de-sorption transition when the binding energy meets
the barrier height $V(z_{\rm br})$. After some algebra, we have
$\sigma_{\rm t}\sim(\epsilon'/\epsilon-1)$ for
$\epsilon'/\epsilon\gg 1$ and $\sigma_{\rm
t}\sim(\epsilon'/\epsilon-1)^3 $ for $\epsilon'/\epsilon\gtrsim
1$. This analytic result is consistent with our simulation data as
shown in Fig.\ref{sigma3.eps}a.

\vspace{15pt}

An approximate solution for the density profile $\rho(z)$ for the
$\sigma>0$ case can be obtained by variational method with trial
wavefunction
\begin{equation}
\psi(z) = \sqrt{\rho_a}(1+\mu\alpha(z-a))
 {\rm e}^{-\frac{1}{2}\alpha(z-a)}
\end{equation}
where $\alpha^{-1}$ is the decay length. $\mu$ is positive because
the trial wavefunction is restricted to be nodeless. $\alpha$ and
$\mu$ are not independent but related via the wavefunction
normalization condition. The inverse decay length is calculated
to be
\begin{equation} \label{decay}
\alpha= 3\beta\Gamma'/2l^2 + \rho_a
\end{equation}
where the leading term is independent of $\sigma$. Near the
transition, the decay length and hence the thickness of the
adsorption layer increase and remain finite. From
Eq.(\ref{rho-sigma}) and (\ref{decay}), we get the scaling
behavior
\begin{equation}
\alpha-\alpha_{\rm t} \sim (\sigma_{\rm t}-\sigma) \qquad
\hbox{for} \quad 0 < \sigma < \sigma_{\rm t} \label{Ds}
\end{equation}
where $\alpha_{\rm t}$ is the threshold inverse layer thickness at
$\sigma=\sigma_{\rm t}^-$. The variation of $\alpha$ as a function
of $\sigma$ obtained from the simulations is shown in
Fig.\ref{sigma3.eps}b which can be well fitted to a linear
relation consistent with Eq.(\ref{Ds}). On the other hand, for the
case of adsorption onto an attractive charged surface ($\sigma<0$)
with substrate of $\epsilon'/\epsilon\leq 1$ (e.g. DNA in aqueous
solution adsorbed onto a charged lipid membrane), the asymptotic
solution to Eq.(\ref{sch}) reproduces the usual scaling
$\alpha\sim|\sigma|^{\frac{1}{3}}$ and is a continuous adsorption
transition \cite{joanny}, and the thickness swells to infinity as
the polyelectrolyte is de-sorbed.



A strongly charged polyelectrolyte immersed in a salt solution
will attract oppositely charged ions to condense until its
effective charge density reaches the Manning threshold
\cite{manning}. This means that one can just renormalize $q_0$ in
our system to $2ea/l_{\rm B}$ if $q_0$ is larger than $2ea/l_{\rm
B}$ ($l_{\rm B}$ is the Bjerrum length). Similarly, the strongly
charged surface of bare charge density larger than $\kappa/(\pi
l_{\rm B})$ is just renormalized back to $\kappa/(\pi l_{\rm B})$
\cite{bocquet}. The Gouy-Chapman length, calculated from the
nonlinear Poisson-Boltzmann theory, is of the order $(l_{\rm
B}\sigma)^{-1}$, which is very large around the transition.
However, the nonlinear Poisson-Boltzmann potential
\cite{israelachvili} near the substrate is given by
\begin{equation}
\phi_{\rm PB}(z)=\frac{2l_{\rm B}}{\epsilon}\ln\frac{1+\gamma{\rm
e}^{-\kappa z}}{1-\gamma{\rm e}^{-\kappa z}}=\phi_0-\frac{2l_{\rm
B}z}{\epsilon\lambda_{\rm GC }}+O(z^2)
\end{equation}
where $\lambda_{\rm GC}$ is the Gouy-Chapman length and
$\gamma=\sqrt{\kappa^2\lambda_{\rm GC}^2+1}-\kappa\lambda_{\rm
GC}$. Notice that the linear term in $z$ is proportional to
$\lambda_{\rm GC}^{-1}\propto\sigma$ and is identical to the
linear term as expanded from the Debye-H\"{u}ckel potential. It is
not surprising since both Poisson-Boltzmann and Debye-H\"{u}ckel
potentials share the same boundary condition. When the adsorbed
polyelectrolyte layer thickness is of one to two monomer size (as
seen from $\alpha^{-1}\sim 1$ in Fig.2b), the surface potential
felt by the polyelectrolyte should be linear. Physically speaking,
it does not matter whether the potential is Poisson-Boltzmann or
Debye-H\"{u}ckel, or even the linear one if the layer thickness is
much smaller than $\lambda_{\rm GC}$ and $\kappa^{-1}$. The
potential near the surface is determined from the boundary
condition. Unlike the case of an attractive surface
\cite{dobrynin}, the effect from large Guoy-Chapman length near
the transition is irrelevant in the high-dielectric case.

Our results on the single polyelectrolyte adsorption may provide a
starting point to study the charge inversion and multi-layer
adsorption \cite{dobrynin2}. At low ionic strength,
polyelectrolytes are adsorbed in a multi-layer structure because
of strong Coulomb repulsion. Each layer is composed of parallel 1d
Wigner crystal \cite{shklovskii}. The upper bound of the
multi-layer thickness is $z_{\rm
br}\sim\sigma^{-\frac{1}{2}}(\epsilon'/\epsilon-1)^{\frac{1}{2}}$.
It suggests we can easily adjust a single layer adsorbed onto a
high-dielectric substrate by tuning the surface charge density.
Rigorous treatment based on this physical picture will be
elaborated elsewhere.

In conclusion, the adsorption transition of a single
polyelectrolyte on a high-dielectric substrate is first order
since the polyelectrolyte needs to overcome a binding energy from
its image charge. Because of the strong Coulomb attraction as
compared to the linear repulsive potential near the surface, the
polyelectrolyte is compressed without any conformational change
before the de-sorption. A scaling law for the adsorption layer
thickness is also derived and verified by simulation.


CHC would like to thank M. Rubinstein for helpful comment. The
work is supported by National Science Council of Republic of China
under Grant No. NSC92-2112-M008-051.

\end{document}